\documentclass[a4paper,11pt]{article}
\pdfoutput=1 
\usepackage{jheppub} 

\usepackage[T1]{fontenc} 

\usepackage[latin9]{inputenc}
\usepackage{float}
\usepackage{amsmath}
\usepackage{amssymb}
\usepackage{graphicx}
 \usepackage{hyperref}
\usepackage{comment}
\usepackage{color}
\usepackage{microtype}
\usepackage{cleveref}
\usepackage{breakurl}
\usepackage{bbm}
\newcommand{\be}{\begin{equation}}
\newcommand{\ee}{\end{equation}}
\newcommand{\ben}{\begin{displaymath}}
\newcommand{\een}{\end{displaymath}}
\newcommand{\bea}{\begin{eqnarray}}
\newcommand{\eea}{\end{eqnarray}}
\def\K{K{\"a}hler}
   \newcommand{\rf}[1]{(\ref{#1})}

\def\be{\begin{equation}}
\def\ee{\end{equation}}
\def\bea{\begin{eqnarray}}
\def\eea{\end{eqnarray}}
\def\ba{\begin{array}}
\def\ea{\end{array}}
\def\bit{\begin{itemize}}
\def\eit{\end{itemize}}

 \makeatletter

\allowdisplaybreaks

\makeatother

\makeatletter
\DeclareRobustCommand{\rcite}[1]{%
  \rcite@aux#1,\@nil{#1}%
}
\def\rcite@aux#1,#2\@nil#3{%
  \if\relax#2\relax
    Ref.~\cite{#3}%
  \else
    Refs.~\cite{#3}%
  \fi
}
\makeatother

\hypersetup{
    colorlinks = true,
    citecolor = {blue},
    linkcolor = {blue},
    urlcolor = {blue},
}

 \title{\rm {\bf \huge KKLT without  AdS}}

\author{ Andrei  Linde}
\affiliation{Stanford Institute for Theoretical Physics and Department of Physics, Stanford University, Stanford, CA 94305, USA}
\emailAdd{alinde@stanford.edu}

\notoc

\abstract{According to the KKLT scenario,  metastable dS vacua are formed as a result of  uplifting of  supersymmetric AdS  vacua  by $\overline {D3}$ branes. I describe an extended version of this scenario where supersymmetric AdS  vacua do not exist, and metastable dS vacua appear after an uplift from a state where the potential of the volume modulus in the absence of $\overline {D3}$ branes would be unbounded below. This mechanism may considerably strengthen vacuum stabilization in the early universe.
}

\begin{document}

\maketitle

   \newpage


 \parskip 4.5pt 

\section{Introduction}
According to the standard version of the KKLT scenario scenario of vacuum stabilization in string theory,   metastable dS vacua are formed due to uplifting of  supersymmetric AdS  vacua by $\overline {D3}$ branes   \cite{Kachru:2003aw}; see  \cite{Douglas:2006es,Akrami:2018ylq} for a general discussion of related issues, and \cite{Hamada:2019ack,Kachru:2019dvo} for some recent progress. 

The 4d supergravity formulation of the KKLT scenario  \cite{Ferrara:2014kva,Kallosh:2014wsa,Bergshoeff:2015jxa} is described by  the superpotential    
\be
W   =W_{0} -Ae^{-aT} + \mu^{2 }X\ .
\label{adssup}
\ee
Here field $T$ is the volume modulus, and $X$ is the nilpotent field $X$ representing the $\overline {D3}$ brane contribution. The  nonperturbative term $ -Ae^{-aT}$ in the superpotential\footnote{Traditionally, the nonperturbative term is written as $Ae^{-aT}$, but we equivalently represent it as $-Ae^{-aT}$, to simplify the description of our main results.}  may appear, for example, in the presence of a stack of D7 branes wrapping a 4-cycle. If there are $N$ branes in the  stack, one has $a =  2\pi/ N$. Alternatively, this term may emerge due to instanton effects. 
 The parameter $A$   depends on  the  values at which the complex structure moduli are stabilized \cite{Burgess:1998jh,Baumann:2006th,Baumann:2010sx}. 
 
If the $\overline {D3}$ brane is in the bulk, which is the main case to be considered in this paper, one can describe  uplifting  by  using the \K\ potential  
\be \label{Xout}
K = -3 \ln (T+\bar T) + X\bar X , 
\ee
 and then taking $X = 0$ after calculating  the potential $V(T)$. 
 Alternatively, if   the anti-D3-brane is in a strongly warped region, its effect can be described by considering the \K\ potential $K = -3 \ln (T+\bar T - X\bar X)$ \cite{Kallosh:2015nia}.
 
 In section \ref{AdS} we will show that supersymmetric AdS vacua do exist in the  KKLT scenario  \cite{Kachru:2003aw}, but only under the condition $0< W_{0}/A < 1$. In section \ref{bottomless} we will find metastable dS vacua obtained by uplifting for $W_{0}/A >1$. In this case  supersymmetric AdS vacua do not exist prior to the uplifting, and  the value of the volume modulus $T$ in the dS vacuum is somewhat smaller than in the standard regime  $0< W_{0}/A < 1$. However, we will show that the stabilized volume modulus $T$ in this model  always remains greater than $a^{{-1}} = {N\over 2\pi}$, even for very large $W_{0}$. In section \ref{disc} we will discuss possible implications of our results in the cosmological context, and show that an increase of $W_{0}$ may significantly strengthen vacuum stabilization in the early universe.

\section{Uplifting from AdS}\label{AdS} 

We consider the  potential $V(T)$ of the field $T$, represent the field $T$ as   $T = t + i \theta$, and search for a minimum of the potential  at $\theta = 0$. One can show that in this theory   $V_{\theta}(\theta = 0) = V_{t,\theta}(\theta = 0) = 0$, and 
\be
V_{\theta,\theta}(\theta = 0)= {a^{3} A e^{-a t}W_{0}\over 2t^{2}}  \ .
\ee
For definiteness, we  consider $ A > 0$. In this case, the state $\theta = 0$ is stable with respect to growth of perturbations of the field $\theta$ for $W_{0} > 0$. 

The potential at $T = t$  prior to the uplifting, i.e. for $\mu = 0$, is given by
\be\label{orig}
V =  {a A e^{-2at}\over 6  t^{2}} \big(A(3+ at)-3e^{at}W_{0}\big) \ .
\ee
Its derivative with respect to $t$ is
\be\label{vmin}
V_{t}=  - {a A e^{-at}\over 6 t^{3}}(2+at) \big(A(3+2 at)-3e^{at}W_{0}\big) \ .
\ee
Comparing   it with the expression for $DW$,
\be\label{adsmin}
DW(t)= {e^{-at}\over 2  t }  \big(A(3+2 at)-3e^{at}W_{0}\big) \ ,
\ee
one finds that any minimum of the potential prior to the uplifting automatically satisfies the condition $DW = 0$, i.e. it is supersymmetric. 
 The potential at the minimum is negative, 
 \be\label{VAdS}
 V_{AdS} = -{a^{2} A^{2}\over 6t}e^{-2at} \ ,
 \ee
  so it is a stable supersymmetric AdS minimum, which is the standard part of the KKLT construction. 

Thus in the KKLT scenario all minima that we can uplift are supersymmetric AdS.  But do we really need to have a stable minimum prior to the uplifting? Naively, the answer is yes, we must have  it, because otherwise what exactly are we going to uplift? However, let us see whether one can relax this requirement.

\section{Uplift from a bottomless well}\label{bottomless}

Equation $DW = 0$ describing the position of the AdS minimum can be represented as follows:
 \be
{W_{0}\over A}=    e^{-x} \Bigl(1+{2\over 3} x\Bigr) \ ,
\ee
where $x = at$. The r.h.s. of this equation increases with the decrease of $x$, and approaches $1$ at $x = 0$. This means that the potential does not have any minimum for $W_{0}/A >1$. 
For $W_{0}/A > 1$  the potential prior to the uplifting is unbounded below, and in the small $t$ limit  it falls to $-\infty$  as 
\be\label{asympt}
V =   -{W_{0}-A\over 2t^{2}} + ... 
\ee

However, this does not mean that this singular potential cannot be stabilized and uplifted. The full expression for the potential, taking into account uplifting, is 
\be\label{uppot}
V =  {\mu^{4}\over 8 t^{3}} +  {a A e^{-2at}\over 6  t^{2}} \big(A(3+ at)-3e^{at}W_{0}\big)\ .
\ee
The first term is a  positive $\overline {D3}$ contribution to $V$ in the theory \rf{adssup},  \rf{Xout}, 
\be\label{upl}
\Delta V = {\mu^{4}\over 8 t^{3}}  \ .
\ee
This contribution immediately makes the potential bounded below. At small $t$ the total potential  
\be
V = {\mu^{4}\over 8 t^{3}}   -{ W_{0}-A\over 2t^{2}} + ... 
\ee
 is dominated by its first, positive term, which stabilizes the potential, and uplifts its minimum.\footnote{This resembles the explanation of stability of the hydrogen atom in quantum mechanics. The  energy of interaction of a proton and an electron is $-{e^{2}\over r}$, which is a bottomless well. But  electron does not fall to $r = 0$ because the energy required for compression of its wave function to $\Delta x \sim r$ is proportional to $1\over r^{2}$.}

To study this effect in a more detailed way, one should find the values of $W_{0}$ and $\mu$ required for the existence of a metastable  dS minimum with a negligibly small positive value (cosmological constant) $V_{dS} = \Lambda \approx 0$ 
at a given point $x = a\, t$  for given $a$, $A$. These  conditions yield:
\be\label{W0}
{W_{0}\over A} ={ e^{-x} (2x^{2} +4x -3)\over 3(x -1)} \ ,
\ee
\be\label{mu}
{ \mu^{4}\over A^{2}} =  {4 x^{2}  e^{-2x} (2  + x)\over 3(x -1)} \ .
\ee
These results imply that for each set of parameters $a$ and $A$ one can find a set of parameters $W_{0}$ and $\mu$ such that the potential $V$ has a metastable dS minimum with a very small positive $V$, at any desirable value of $x = at >1$, i.e. at $T=t > 1/a$.  

Equations \rf{W0} and \rf{mu} are equally valid for $W_{0}/A >1$ and for $W_{0}/A < 1$. From this perspective, there is nothing special about  the dS states uplifted from the supersymmetric AdS minima with $W_{0}/A < 1$, as compared to the dS states which appear after adding the  positive $\overline {D3}$ contribution \rf{upl} to the potential \rf{orig}, which is unbounded below at small $t$ for $W_{0}/A >1$.

For $W_{0} <A$ one has  $at > 1.569$, and the potential prior to the uplifting has a supersymmetric AdS minimum, which is then uplifted to dS. On the other hand, all dS vacua with a minimum at $1< at < 1.569$ are obtained by adding the uplifting $\overline {D3}$ contribution \rf{upl} to the   potential \rf{orig}, which is unbounded below for $W_{0} > A$ in the absence of the positive $\overline {D3}$ contribution  \rf{upl}. Note that the position of the minimum always remains at $t > 1/a$, and it gradually approaches $1/a$ only in the limit $W_{0}\to \infty$.

In the limit $ at-1\ll 1$ one has
\be
at -1=  {  e^{-1} A \over W_{0}  } =
 {4    e^{-2 } A^{2} \over   \mu^{4} } \ .
\ee
Therefore uplifting to a dS state with a tiny positive cosmological constant and  $W_{0} \gg A$  requires uplift with 
\be
\mu^{4}={4\over e} A W_{0} \ .
\ee
The moduli mass squared of the field $t$ after the uplift with $W_{0} \gg A$ is
\be
m_{t}^{2}= {a ^{3}\over 3e} \, A W_{0}  \ ,
\ee
and the gravitino mass   is
\be
m_{3/2}= \left({a\over2}\right)^{3/2}  W_{0}  \ .
\ee

\begin{figure}[H]
\center
\includegraphics[scale=0.52]{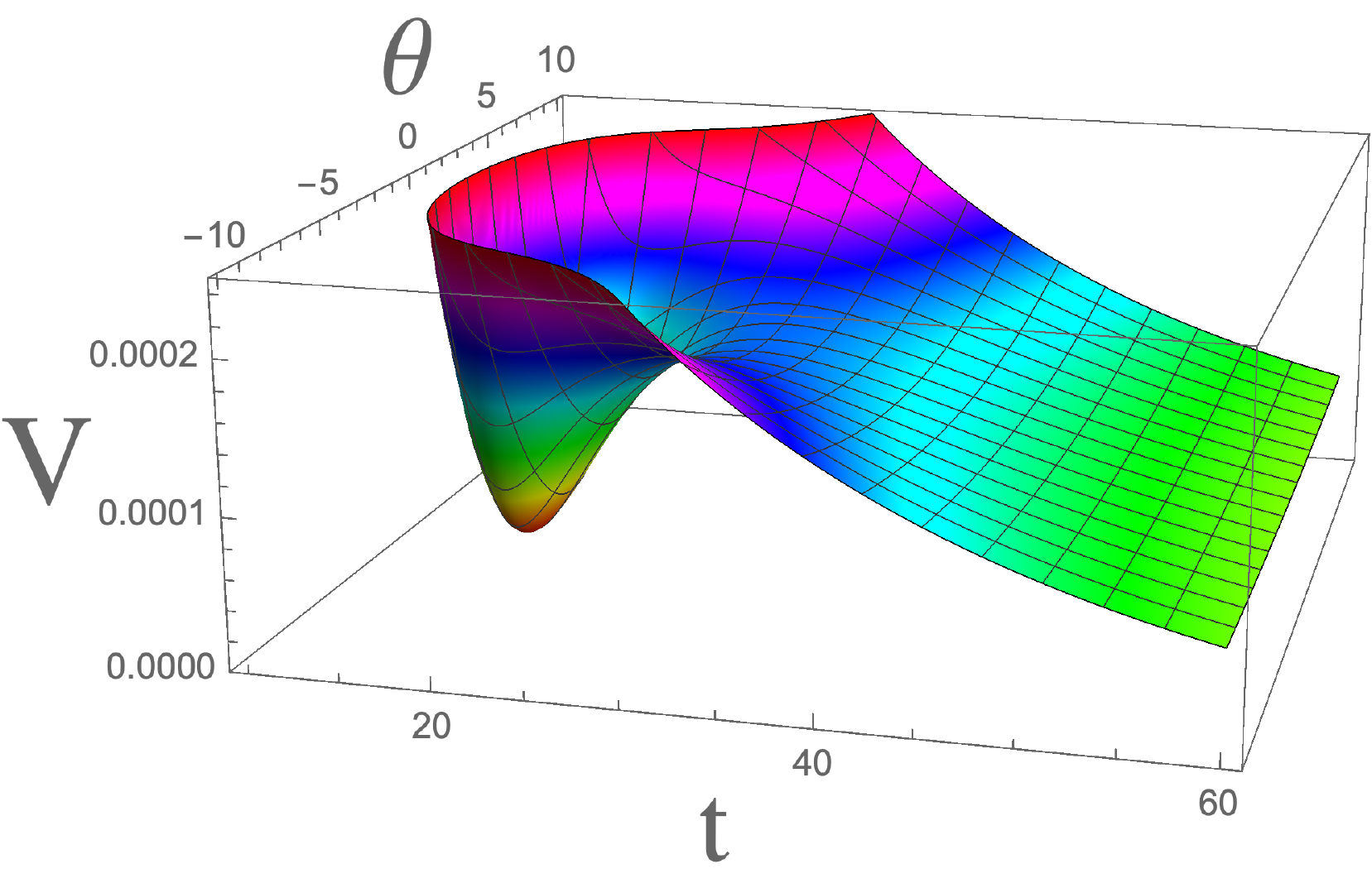}
\caption{\footnotesize KKLT potential \rf{uppot} with a metastable dS vacuum at $t \approx 16$, $\theta = 0$. It was obtained by adding the $\overline {D3}$ contribution \rf{upl} with $\mu = 3.4786$  to the  potential   \rf{orig}   in the KKLT model  \rf{adssup}, \rf{Xout} with $a = 2\pi/100$, $A = 1$, $ W_{0} = 100$. }
\label{fig:1}
\end{figure}

Fig. \ref{fig:1} shows the potential $V(T)$ with a dS minimum with a tiny cosmological constant for a particular case $A = 1$, $ W_{0} = 100$, $a = 2\pi/100$. The minimum is  very close to the limiting value of $T=t =1/a \approx15.9$.

\section{Discussion}\label{disc}

In this paper we have shown that   the existence of a supersymmetric AdS (or Minkowski) vacuum is {\it not} a necessary precondition for the existence of stable dS vacua in the KKLT scenario. In particular, when finding dS vacua  using equations \rf{W0}, \rf{mu} we did not make any assumptions about the behavior of the potential in the absence of the uplifting contribution of the $\overline {D3}$ brane. Instead of that, we simply analyzed the behavior of the system for all possible relations between the model parameters, taking into account uplifting.
If, instead, we would divide the procedure into two parts, first finding a supersymmetric AdS vacuum, and then uplifting it, then we would find dS vacua only for $W_{0} < A$  and  miss all strongly stabilized dS states with $W_{0} > A$.

 Thus  the more general approach described above significantly increases the phase space of the possible KKLT parameters.\footnote{Note that in this paper we investigate the original KKLT model   \cite{Kachru:2003aw}, with the uplift interpreted in terms of the nilpotent fields   \cite{Ferrara:2014kva,Kallosh:2014wsa,Bergshoeff:2015jxa}. Other possibilities to have vacuum stabilization with large $W_{0}$ emerge, for example, if one qualitatively changes the structure of the  scalar potential in the KKLT scenario by  adding $\alpha'$ corrections to the \K\ potential, 
as well as non-perturbative contributions to the superpotential, as in the theory of Large Volume Stabilization (LVS)  \cite{Balasubramanian:2004uy,Balasubramanian:2005zx}.}
We found metastable dS vacua not only in the previously considered models with $W_{0} < A$ and uplift from AdS, but also in the models with $W_{0} > A$, where no supersymmetric AdS vacua are available. One can show that the value of the volume modulus $T$ at the minimum of the potential does not change if one simultaneously rescales $W_{0} \to c W_{0}$, $A\to c A $ and $\mu^{2} \to c \mu^{2}$.  Thus this scenario allows considerable freedom in the choice of the parameters.  However, one should check whether there are some constraints following from string theory and/or phenomenology, which would require $W_{0} < A$.

Historically, there were several opposite arguments concerning the value of $W_{0}$ in the KKLT scenario. One of the arguments was that in order to have low-scale supersymmetry breaking one would need to have an extremely small  value of $|W_{0}|$. 
 However, a subsequent investigation has shown that the universe in the KKLT scenario tends to decompactify for the Hubble constant greater than the gravitino mass, which resulted in the stability constraint $H \lesssim m_{{3/2}}< W_{0}$ \cite{Kallosh:2004yh,Buchmuller:2004tz}. A similar constraint in the LVS models is even stronger, $H \lesssim m_{{3/2}}^{3/2}$ \cite{Conlon:2008cj}. Thus the requirement of vacuum stability in the very early universe, where the Hubble constant $H$ was extremely large, tends to disfavor small values of $W_{0}$ in the KKLT construction. 
 
 The simplest way to avoid vacuum destabilization in the early universe compatible with a low-scale supersymmetry is to consider the KL generalization of the KKLT construction \cite{Kallosh:2004yh,BlancoPillado:2005fn,Kallosh:2011qk,Linde:2011ja},  disentangling the strength of the vacuum stabilization and the magnitude of supersymmetry breaking.   This approach  turned out very helpful for dS vacua stabilization  in a broad class of type IIB and type IIA string theory inspired models, and in M-theory \cite{Kallosh:2019zgd,Cribiori:2019drf,Cribiori:2019hrb}. In such models, the smallness of $ W_{0} $ is not required for the smallness of supersymmetry breaking. 
 
The second argument in favor of models with $W_{0} < A$ is that  large values of the volume modulus $T$ are required for suppression of $\alpha'$ corrections and validity of the effective supergravity approach. However, as we already mentioned, the value of the volume modulus at the minimum of the KKLT potential in the model \rf{adssup}, \rf{Xout} is always greater than $1/a = N/2\pi$. Thus the requiredment $T\gg  1$ is automatically satisfied, for any $W_{0}$, if  ${N\over 2\pi} \gg 1$. 

Admittedly, the requirement ${N\over 2\pi} \gg 1$ is a significant constraint on the model. The idea of the more traditional approach with $W_{0}< A$ was to make $T$ much greater than $1/a= N/2\pi$ by considering extremely small $W_{0}$. But the volume modulus depends on $W_{0}$ only logarithmically. Using equation \rf{W0} one can show that if we want to increase the volume modulus 10 times, making it $10/a$ instead of $1/a$, or reduce the required number $N$ ten times while keeping $T$ unchanged, we would need to take $W_{0}/A < 4\times 10^{{-4}}$, and if we want to increase the volume modulus 20 times, we would need to take  $W_{0}/A < 3\times 10^{{-8}}$.  

The models  with $W_{0}\ll 1$ do exist, but constructing such models turned out to be quite difficult.  An example of the model with $W_{0} \sim 10^{-8}$ was only recently found in  \cite{Demirtas:2019sip}. Until we have a much better grasp of the theory, it is hard to tell whether it is easier to construct models with large $N$, or with extremely small $W_{0}$. Meanwhile, as we already mentioned, small values of $W_{0}/A$ in combination with large values of the volume modulus  make dS vacua  vulnerable with respect to decompactification in the early universe   \cite{Kallosh:2004yh,Buchmuller:2004tz}. Therefore the model with  $W_{0}> A$ may have a substantial advantage in describing the early stages of the evolution of the universe. 

\begin{figure}[h!]
\begin{center}
\includegraphics[scale=0.5]{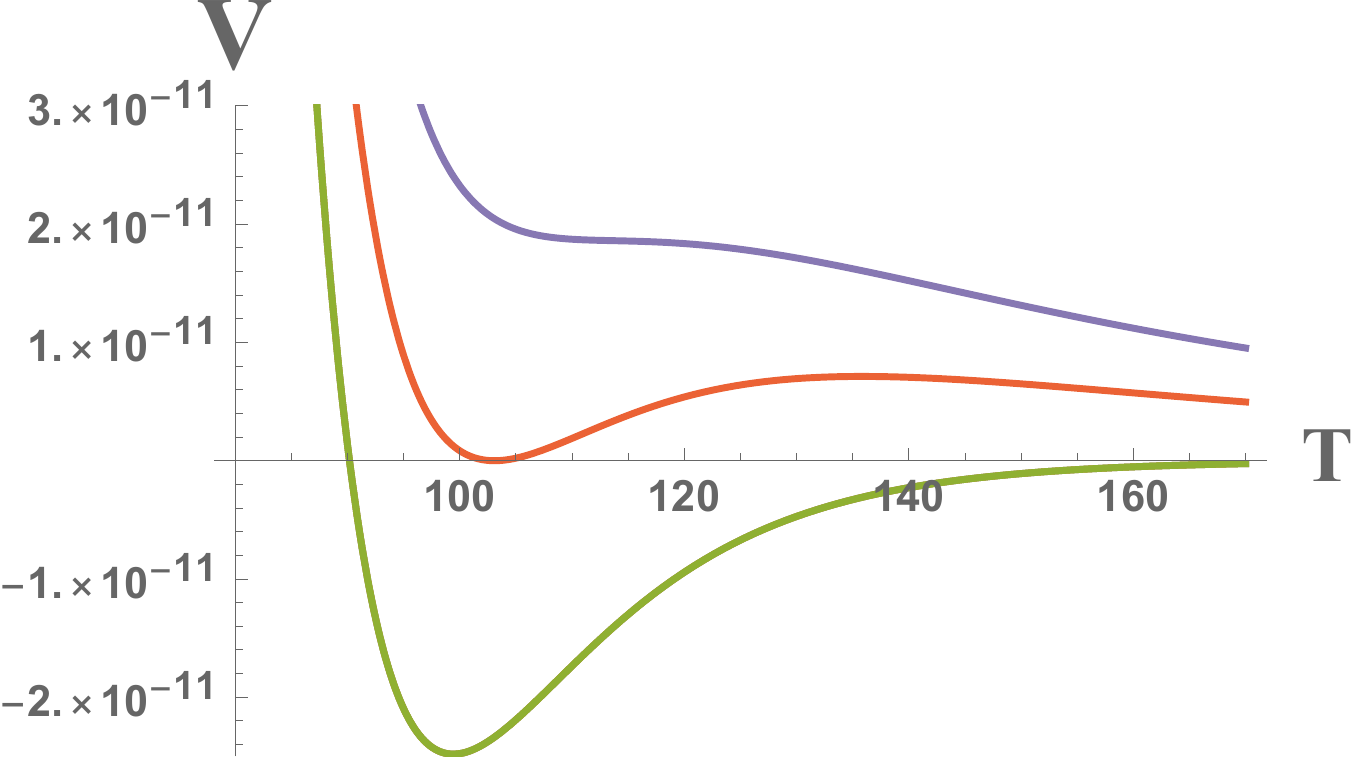}  \qquad
\includegraphics[scale=0.48]{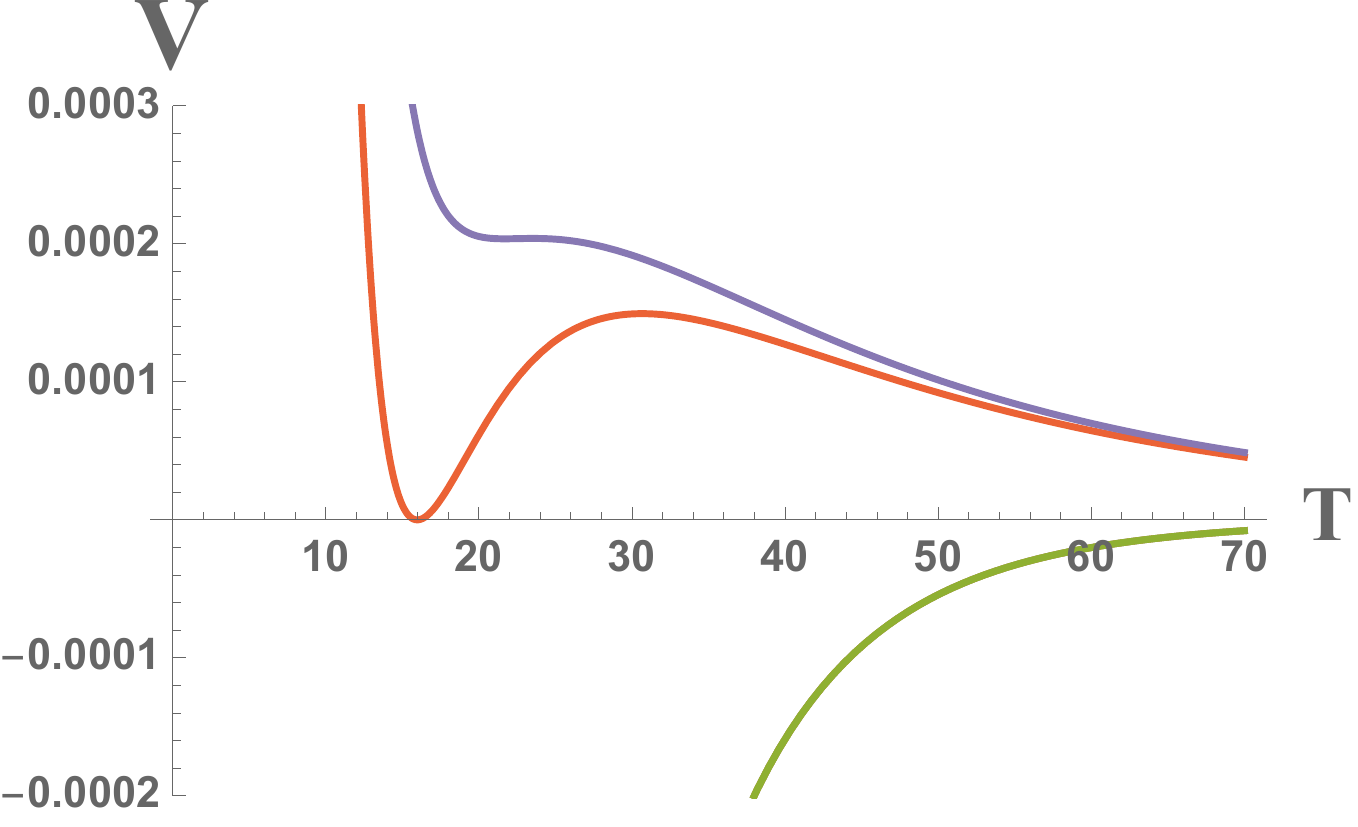}  
\end{center}
\vskip -0.5cm 
\caption{\footnotesize  A comparison of KKLT uplifting for $A = 1$, $a = 2\pi/100$, for  $W_{0} = 10^{{-2}}$ (left panel) and  $W_{0} = 10^{{2}}$ (right panel). The green lines show the potential before the uplifting, the red lines show the potential with a dS minimum with a tiny positive cosmological constant. The blue lines illustrate the disappearance of the minimum if the uplifting is too large, which may result in   decompactification of 6 extra dimensions in the early universe. }
\label{f2}
\end{figure}

To illustrate this point, we show the KKLT potential $V(t)$  for $\theta = 0$, $A = 1$ and  $a = 2\pi/100$, for two very different values of  $W_{0}$: $W_{0} = 10^{{-2}}$ (left panel of Fig. \ref{f2}) and  $W_{0} = 10^{{2}}$ (right panel of Fig. \ref{f2}).

In the first case, for $W_{0 } =  10^{{-2}}$, uplifting occurs from a supersymmetric AdS vacuum (the green line). The dS minimum of the uplifted potential shown by the red line is at $t \sim 103$. The barrier stabilizing the vacuum state with a small cosmological constant has the height  $\sim 10^{{-11}}$ in  the Planck density units. If one considers a more significant uplifting, which is similar to what may happen in the early universe at large energy density  \cite{Kallosh:2004yh,Buchmuller:2004tz}, the dS minimum disappears and the universe decompactifies. This happens for energy density greater than $2 \times 10^{{-11}}$, which is an order of magnitude below the energy density during inflation in many popular inflationary models, such as the Starobinsky model, the Higgs inflation, and the simplest versions of $\alpha$-attractors.  

 The right panel  (see also Fig. \ref{fig:1}) shows the model with  $A = 1$, $ W_{0} = 100$, which is unbounded below prior to the uplifting. In this case the uplifted potential shown by the red line has a minimum at $t \approx 16$, which is close to the limiting value of $t =1/a =100/2\pi \approx 15.9$. The stabilizing barrier is 9 orders of magnitude higher than in the case  $W_{0 } =  10^{{-2}}$, and the dS vacuum is stable in the early universe at energy density up to $2\times 10^{-4}$. 
 
To understand the general pattern revealed by these two figures, let us study the standard regime with $W_{0} \ll A$ and AdS minimum with the depth given by \rf{VAdS}. Its uplifting is achieved by adding to the potential a function rapidly decreasing at large $t$. Therefore the height of the barrier stabilizing the dS minimum after the uplifting is always smaller than the depth of the AdS minimum \cite{Kallosh:2004yh}, 
\be
\label{barrier}
 V_{\rm barrier} < {a^{2} A^{2}\over 6t}e^{-2at}  \ .
 \ee
If $W_{0}$ is small and the minimum of the potential is at $t \gg 1/a$, then this expression  in combination with \rf{vmin} implies that 
\be
\label{barrier2}
 V_{\rm barrier} < {3 W_{0}^{2} \over 8t^{3} } \ .
 \ee
The  maximal height of uplifting of a dS vacuum shown by the inflection point of the blue lines in the left panel Fig. \ref{f2} is just a little bit higher than $ V_{\rm barrier}$. Thus the traditional approach results in the suppression of the height of the protective barrier by two factors: by $W_{0}^{2}$, which was supposed to be very small, and by  $t^{-3}$, where $t$ was supposed to be very large.

On the other hand, in the limit $W_{0} \gg A$, instead of the calculating the height of the barrier one can obtain a direct analytical estimate of the maximal height of uplifting of a dS vacuum,  shown by the inflection point in the right panel of Fig. \ref{f2}. The inflection point appears at $t \approx \sqrt{2}/a$, and its height is
\be\label{infllim}
V_{\rm max}  \approx {\sqrt{2} - 1\over 12}e^{-\sqrt 2} a^{3} A\, W_{0}  \sim 10^{{-2}}  a^{3} A\, W_{0}\ .
\ee

In the example shown in Fig. \ref{f2} we see that one can increase the range of stability of compactification by 9 orders of magnitude while still having a reasonably large value of the volume modulus. Equation \rf{infllim} shows that one can further enhance  stability of compactification by increasing $A$ and $W_{0}$ while preserving the same value of the volume modulus $T$, which does not depend on $W_{0}$ in the regime when $W_{0} \gg A$. 

These results have been obtained in the version of the KKLT scenario  with the \K\ potential $K = -3 \ln (T+\bar T) + X\bar X$ describing uplift due to the $\overline {D3}$ brane in the bulk. If one considers uplift due to the $\overline {D3}$ brane in the strongly warped region, using the \K\ potential $K = -3 \ln (T+\bar T- X\bar X)$, the results change. Stable dS vacua may still exist for  $W_{0}> A$ remains possible, but only for a rather limited range of values of $W_{0}/A$. This suggests that it is easier to achieve strong vacuum stabilization in the models where the $\overline {D3}$ brane is in the bulk.

Similar results can be obtained in a more general class of theories. In particular, one may consider the KL version of the KKLT scenario, which  allows to have small supersymmetry breaking compatible with  strong moduli stabilization \cite{Kallosh:2004yh,BlancoPillado:2005fn,Kallosh:2011qk,Linde:2011ja}. The basic idea was to find a supersymmetric Minkowski vacuum without flat directions. Any small deformation of such vacuum due to a change of model parameters transforms it into a supersymmetric AdS vacuum, which can be subsequently uplifted. If the deformations of the original state are sufficiently small, one obtains a strongly stabilized dS vacuum with a controllably small supersymmetry breaking. 

In a recent series of papers  \cite{Kallosh:2019zgd,Cribiori:2019drf,Cribiori:2019hrb} this approach was generalized and used for finding stable dS vacua in a broad class of type IIB and type IIA string theory  models, and in M-theory, with many moduli. In addition to many  dS vacua obtained by small deformations of the original supersymmetric Minkowski vacua,  Ref.  \cite{Kallosh:2019zgd,Cribiori:2019drf,Cribiori:2019hrb} also  found stable dS vacua produced by a very large increase of $W_{0}$ accompanied by a large uplift. General  theorems describing small deformations of a supersymmetric Minkowski vacuum state \cite{BlancoPillado:2005fn,Kallosh:2019zgd,Cribiori:2019drf} did not make any predictions about existence and stability of vacua after a very large increase of $W_{0}$, and yet such dS vacua were found. Moreover, they were stabilized by potential  barriers which could be many orders of magnitude higher than the barriers stabilizing the original Minkowski vacua. 
In particular, one can show that the strongly stabilized dS vacuum in the KL scenario with the racetrack superpotentials shown in Fig. 15 of \cite{Cribiori:2019drf} have been obtained  in the model, which, in the absence of the uplifting contribution, would have a potential $V(T)$ unbounded from below, just as in the KKLT scenario  with $W_{0}> A$.  
\vskip 2pt

To summarize, the results presented in this paper  show that there are two different ways to construct metastable dS vacua  in the KKLT model \rf{adssup} \rf{Xout}. The standard approach based on the uplifting of supersymmetric AdS vacua is applicable only for $W_{0} <A$ in the KKLT superpotential  \rf{adssup}. However, we found that  metastable dS states can be constructed for $W_{0} > A$ as well, despite the absence of  supersymmetric AdS vacua in that regime. The most important property of the new dS stabilization regime with $W_{0} > A$  is that it allows to achieve a much stronger vacuum stabilization than in the standard version KKLT scenario with $W_{0} <A$.  This property may be helpful for constructing consistent cosmological models based on string theory.  
  
 \vskip 5pt 

 The author is grateful to Renata Kallosh for many enlightening discussions, and to N.~Cribiori, S. Kachru, L. McAllister, C. Roupec and Y. Yamada for useful comments. I am supported by SITP and by the US National Science Foundation Grant  PHY-1720397, and by the  Simons Foundation Origins of the Universe program (Modern Inflationary Cosmology collaboration),  and by the Simons Fellowship in Theoretical Physics.

 \bibliographystyle{JHEP}
\bibliography{lindekalloshrefs}
\end{document}